\documentclass[onecolumn,amsmath,amssymb,superscriptaddress]{revtex4}
\usepackage{amsfonts,amssymb,bm,graphicx,times} \usepackage{amsmath}
\usepackage{float,color}

\begin{document}
\title{Quantum electrodynamics of one-photon wave packets}

\author{M.Stobi\'nska$^{(1)}$, G. Alber$^{(2)}$,
  G. Leuchs$^{(1)}$\\ $^{(1)}$~{\em Max Plack Institute for the
    Science of Light - G\"unther-Scharowsky-Str.1, Bau 24, 91058
    Erlangen, Germany}\\ $^{(2)}$~{\em Institut f\"ur Angewandte
    Physik, Technische Universit\"at Darmstadt, Hochschulstra\ss e 4a,
    64289 Darmstadt, Germany}}
\maketitle

\tableofcontents
\section{Introduction}

Excited bound states of atoms are one of the simplest examples of
unstable quantum states which decay radiatively into the continuum of
possible one-photon states of the electromagnetic field in free
space. The search for a satisfactory theoretical explanation of the
line frequencies and intensities of photon emission spectra at the
beginning of the last century eventually culminated in Heisenberg's
'magical paper' from July 1925 \cite{Heisenberg,Aitchison} and in the
development of modern quantum mechanics
\cite{Born,Schroedinger1,Schroedinger2,Dirac}. Although basic aspects
of this spontaneous decay process have already been described
theoretically in an adequate way in the early days of modern quantum
mechanics \cite{WignerWeisskopf1,WignerWeisskopf2}, interestingly,
some of its time-dependent dynamical aspects are still of topical
current interest.

To a large part the current interest in this elementary photon
emission process is due to recent advances in quantum technology. They
have enabled one to prepare quantum states of matter and of the
electromagnetic field to such a high level of precision that not only
subtle quantum electrodynamical phenomena can be tested experimentally
but that these phenomena can also be exploited practically for
purposes of quantum information processing \cite{quantuminfoproc}.  On
the one hand it is possible to engineer quantum states of individual
atoms or even of multi-particle systems with the help of sophisticated
particle traps. In this context electromagnetic fields play a
predominant role for achieving the required trapping and they also
allow to control the material quantum states involved with an
unprecedented level of accuracy \cite{trappedatoms1,trappedatoms2}.
On the other hand it is also possible to engineer quantum states of
the electromagnetic field.  Not only classical electromagnetic fields
with large photon numbers can be generated by sophisticated pulse
shaping techniques \cite{pulseshape} but also particular few-photon
quantum states of the electromagnetic field can be prepared in a
controlled way.  In the context of cavity quantum electrodynamics
\cite{CQED}, for example, extreme selection of electromagnetic field
modes by cavities in combination with trapping of single atoms even
enables one to control the interaction between single atoms and the
electromagnetic field.  Thus, it is possible to prepare single-photon
single-mode quantum states of the electromagnetic field
\cite{onephoton}, for example, or to excite a single atom by a
single-photon quantum state perfectly with the help of vacuum Rabi
oscillations \cite{Schleich}.  Despite these significant advances of
quantum electrodynamics similarly efficient techniques for preparing
many-mode few-photon quantum states of the electromagnetic field or
for controlling the resulting atom-field dynamics are significantly
less well developed. Advances in this direction would be particularly
interesting for purposes of quantum information processing.  They
would significantly enhance the flexibility of exchanging quantum
information between the electromagnetic field, which is particularly
useful for the transport of quantum information, and matter, which is
well suited for the storage of quantum information.

In this contribution we investigate quantum electrodynamical many-mode
aspects by exploring the simplest possible situation in this context,
namely the interaction of a single atom, modeled by a simple two-level
system, with many-mode one-photon quantum states of the
electromagnetic field \cite{Quabis}.  In particular, we concentrate on
the question how the engineering of electromagnetic field modes
influences the atom-field interaction. An interesting problem in this
context is the engineering of ideal one-photon multi-mode quantum
states which may excite a single atom perfectly \cite{StobAlLeu}. For
purposes of quantum information processing such one-photon multi-mode
states are particularly well suited for transmitting quantum
information and for storing it in a material memory again.

This article is organized as follows: In Sec. \ref{II} characteristic
aspects of two extreme cases of the interaction of a simple two-state
quantum system with the radiation field are discussed, namely the
single-mode Jaynes-Cummings-Paul model \cite{JCP1,JCP2} and
spontaneous photon emission in free space
\cite{WignerWeisskopf1,WignerWeisskopf2}.  In Sec. \ref{III} we
explore dynamical modifications originating from modifications of the
electromagnetic field modes involved. For this purpose the interaction
of a single photon with a two-level system in a closed spherical
cavity of arbitrary size \cite{Alber1992} and in a half-open parabolic
cavity \cite{Sondermann2007,Maiwald} are explored.

\section{Quantum electrodynamics of a material two-level system - basic aspects\label{II}}

In this section basic results concerning the interaction of matter with the quantized radiation field are discussed.
For this purpose an elementary two-level model of matter is considered with involves two non-degenerate relevant energy eigenstates which are coupled almost resonantly to the electromagnetic field. Characteristic quantum electrodynamical properties of this model system are particularly apparent in cases in which the
quantum states of the electromagnetic field contain only a small number of photons.
The resulting dynamics depends significantly on whether only one mode of the radiation field or an infinite
number of them participate in the interaction.

\subsection{The Jaynes-Cummings-Paul model}
One of the simplest models which describes characteristic quantum features of the almost resonant interaction of matter with the radiation field is the Jaynes-Cummings-Paul model \cite{JCP1,JCP2}. In this model a material two-level system interacts with a single mode of the radiation field. In the Schr\"odinger picture its dynamics is described by the Hamiltonian
\begin{eqnarray}
\hat{H} &=& E_g|g\rangle \langle g| + E_e |e\rangle \langle e| +  \hbar \omega \hat{a}^{\dagger} \hat{a} + \hbar g \hat{a}|e\rangle \langle g| + \hbar g^* \hat{a}^{\dagger}|g\rangle \langle e|.
\end{eqnarray}
The energies of the material two-level system are denoted by $E_g$ and $E_e > E_g$
and $\hat{a}$ ($\hat{a}^{\dagger}$) is the destruction (creation) operator of the almost resonantly coupled electromagnetic field mode of frequency $\omega$, i.e. $E_e - E_g \approx \hbar \omega$. The coupling constant $g$ characterizes the strength of the interaction of the material two-level system with the single mode of the radiation field. In the dipole approximation, for example, which applies in typical quantum optical situations and in which it is assumed that
the wavelength of the almost resonantly coupled electromagnetic field mode is significantly larger than the extension of the material charge distribution of the two-level system, this coupling constant is given by
\begin{eqnarray}
\hbar g \hat{a} &=& - \langle e|\hat{{\bf d}}|g\rangle \cdot \hat{{\bf E}}_{+}({\bf x}_0).
\end{eqnarray}
Thereby, $\hat{{\bf d}}$ is the material dipole operator.  The positive frequency component of the electric field operator is denoted
by
\begin{eqnarray}
\hat{{\bf E}}_{+}({\bf x}) &=& i\sqrt{\frac{\hbar\omega}{2\epsilon_0}} {\bf u}({\bf x})\hat{a},
\end{eqnarray}
and it has to fulfill the transversality condition $\nabla \cdot \hat{{\bf E}}_{+}({\bf x}) =0$. (
$\epsilon_0$  
is the permittivity of the vacuum.)
The normalized mode function ${\bf u}({\bf x})$ is a solution of the Helmholtz equation
\begin{eqnarray}
\left(\nabla^2 + \omega/c^2\right){\bf u}({\bf x})  &=&0
\label{Helmholtz}
\end{eqnarray}
and fulfills the boundary conditions of the mode selecting cavity involved. It is normalized according to the relation
$\int_{\mathbb R^3} d^3 x~|{\bf u}({\bf x})|^2 =1$. The position of the center-of-mass of the material charge distribution is denoted by
${\bf x}_0$. 

Within this model the dynamics of the almost resonantly coupled matter-field system can be described in a straight forward way by expanding the
quantum state $|\psi\rangle_t$ at time $t$ in the basis of energy eigenstates of the uncoupled quantum system, i.e.
\begin{eqnarray}
|\psi\rangle_t &=& \sum_{n=0}^{\infty} \{ a_{e,n}(t) e^{-i(E_e + n\hbar\omega)t/\hbar}|e\rangle\otimes |n\rangle + a_{g,n}(t) e^{-i(E_g + n\hbar\omega)t/\hbar}|g\rangle\otimes |n\rangle\}.
\end{eqnarray}
The time-dependent Schr\"odinger equation $i\hbar d|\psi\rangle_t/dt = \hat{H}|\psi\rangle_t$ yields the system of pairs of coupled equations
\begin{eqnarray}
\dot{a}_{e,n}(t) &=& -ig\sqrt{n+1}e^{i\Delta t}a_{g,n+1}(t),\nonumber\\
\dot{a}_{g,n+1}(t) &=& -ig^*\sqrt{n+1}e^{-i\Delta t}a_{e,n}(t)~~({\rm for} ~~n \geq -1)
\label{set}
\end{eqnarray}
with $\Delta = (E_e - \hbar \omega - E_g)/\hbar$ denoting the detuning from resonance.

If initially, at $t=0$, the two-level system is prepared in its excited state $|e\rangle$, i.e. $a_{g,n}(0)=0$ for $n\geq 0$,
Eqs.(\ref{set}) yield the solution
\begin{eqnarray}
a_{e,n}(t) &=&a_{e,n}(0)\left( \cos\left(\frac{\Omega_nt}{2}\right) - \frac{i\Delta}{\Omega_n}\sin\left(\frac{\Omega_nt}{2}\right)\right)e^{i\Delta t/2},\nonumber\\
a_{g,n+1}(t) &=-&a_{e,n}(0) \frac{2ig^*\sqrt{n+1}}{\Omega_n}\sin\left(\frac{\Omega_nt}{2}\right)e^{-i\Delta t/2}.
\end{eqnarray}
The time evolution of the probability amplitudes $a_{e,n}(t)$ and $a_{g,n}(t)$ exhibits a characteristic periodic energy transfer between the two-level system and the electromagnetic field mode which is characterized by the
$n$-photon Rabi frequency $\Omega_n = \sqrt{\Delta^2 + 4|g|^2 (n+1)}$.
As the period of this energy exchange depends on the photon number the resulting time evolution exhibits interesting
collapse and revival phenomena.

\begin{figure}
\begin{center}
      \scalebox{0.5}{\includegraphics{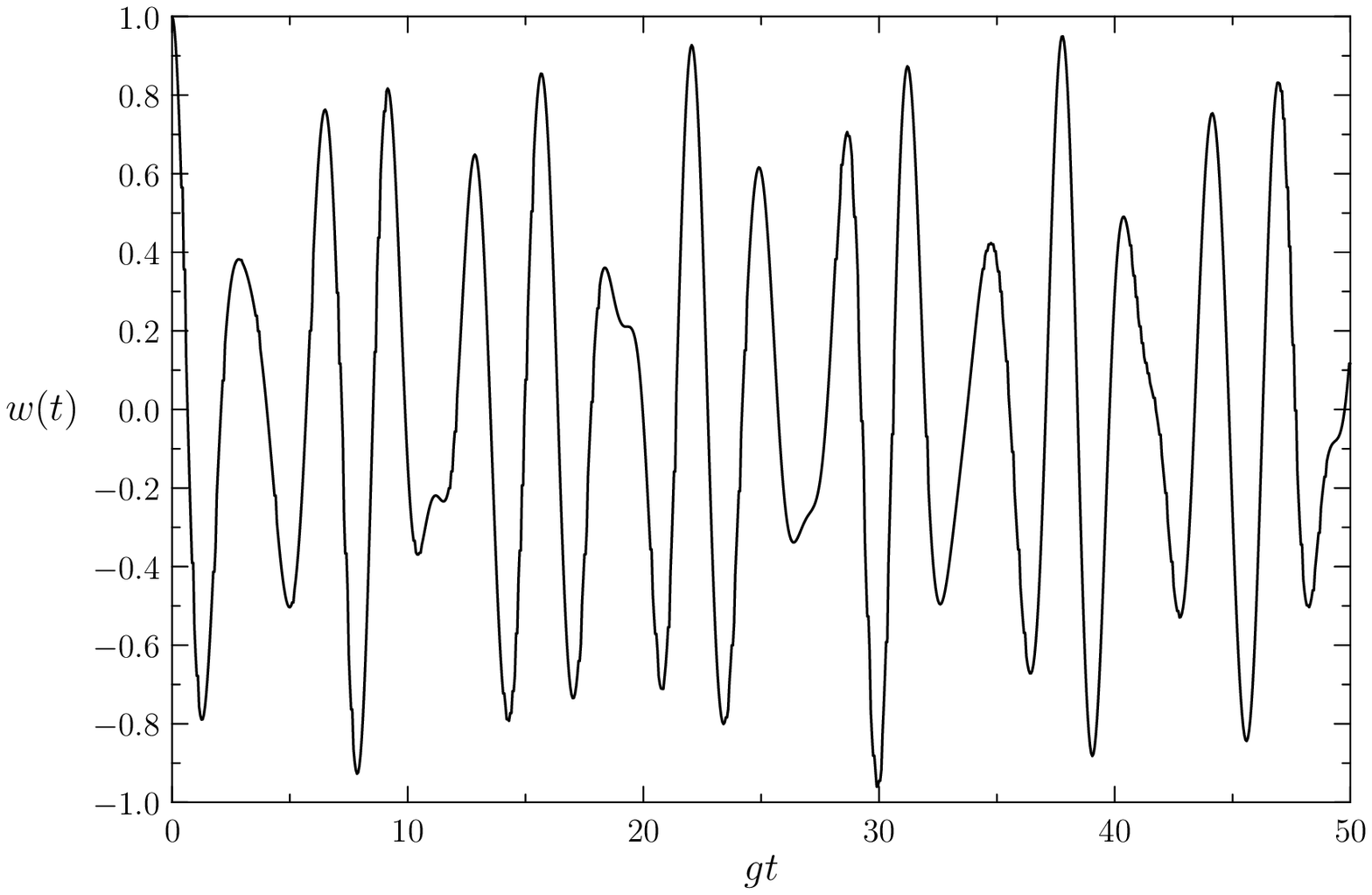}}
     \scalebox{0.5}{\includegraphics{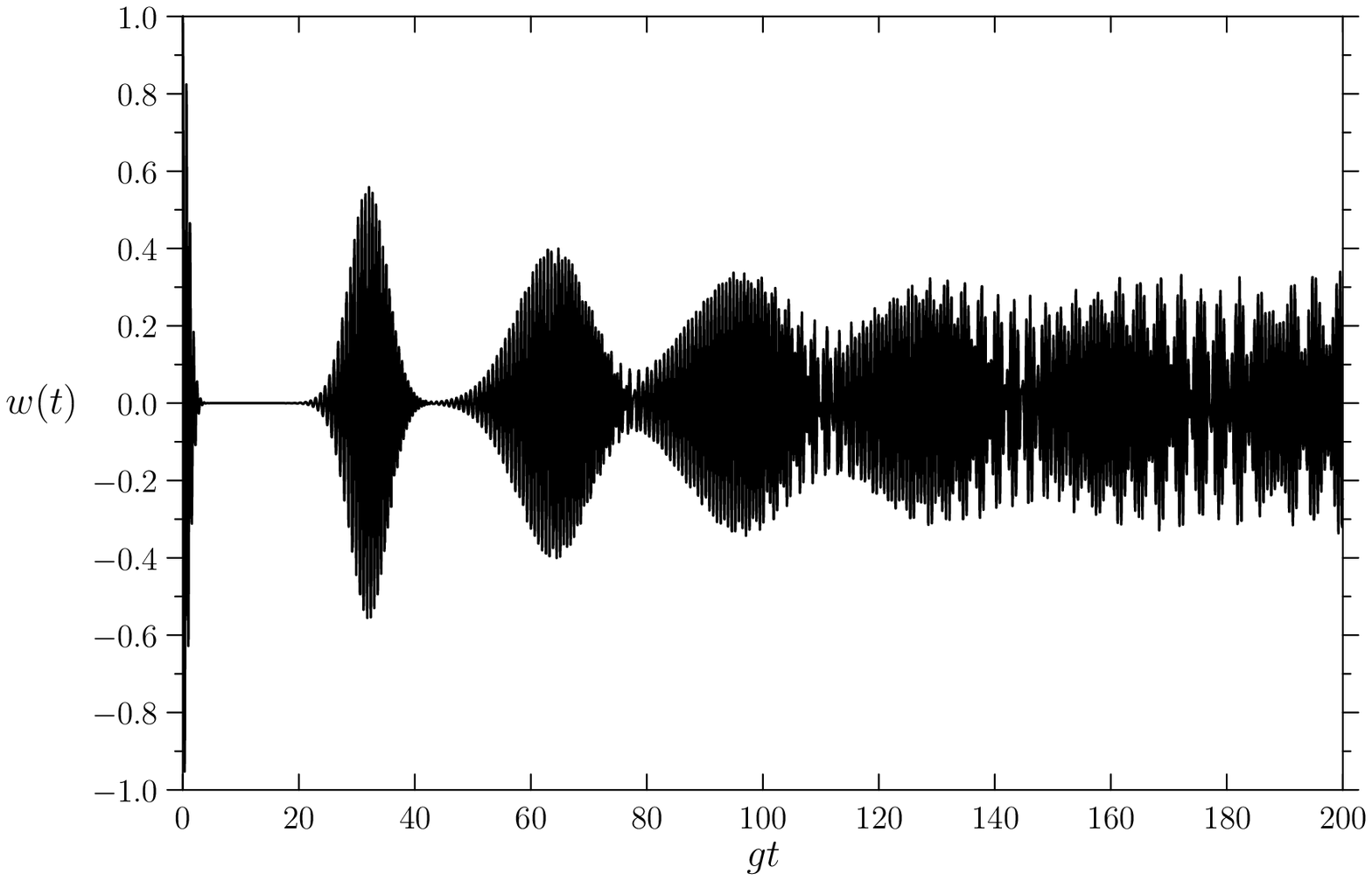}}
  \end{center}
  \caption{Time dependence of the inversion $w(t)$ of Eq.(\ref{inversion}) of the two-level system describing its excitation by
  a coherent single-mode state $|\alpha\rangle$: $<n> = 0.5$ (left curve), $<n> = 25$ (right curve).}
  \label{collapse}
\end{figure}

Let us consider a special case in more detail
in which initially the two-level system is excited and the single-mode radiation field is in a coherent state $|\alpha\rangle$ with
$\hat{a}|\alpha\rangle = \alpha |\alpha\rangle$ and $\alpha \in {\mathbb C}$ \cite{Schleich}.
The resulting time evolution of the inversion of the two-level system is given by
\begin{eqnarray}
w(t) &:=& \sum_{n=0}^{\infty} \{|a_{e,n}(t)|^2 - |a_{g,n}(t)|^2\} =
\sum_{n=0}^{\infty}|a_{e,n}(0)|^2 \left( \frac{\Delta^2}{\Omega_n^2} + \frac{4|g|^2(n+1)}{\Omega_n^2}\cos\left(\Omega_n t\right) \right)\nonumber\\
&&
\label{inversion}
\end{eqnarray}
with $a_{e,n}(0) = {\rm exp}(-|\alpha|^2/2)\alpha^n/\sqrt{n!}$. In Fig.(\ref{collapse}) the time evolution of this inversion is depicted for resonant coupling, i.e. $\Delta =0$, and various values of the mean photon number 
$<n> = \langle \alpha |\hat{a}^{\dagger}\hat{a}|\alpha\rangle = |\alpha|^2$ of a coherent state $|\alpha\rangle$. It is apparent that the initially prepared inversion collapses after a 
characteristic collapse time of the order of $T_{c} = 2\pi/|g|$. 
Furthermore, at multiples of times of the order of $T_{r} = 2\pi \sqrt{<n>+1}/|g|$
the probability amplitudes interfere constructively again thus causing revivals of the initial excitation \cite{revivals1,revivals2}.
Finally,
it should also be mentioned that according to Eq.(\ref{inversion}) the appearance of collapse and revival phenomena do not necessarily require an initially prepared pure quantum state of the electromagnetic field.

If one considers the extreme case that the two-level system is excited initially but the electromagnetic field is in its ground (vacuum) state one obtains the result
\begin{eqnarray}
w(t) &=& \frac{\Delta^2}{\Omega_0^2}+\frac{4|g|^2}{\Omega_0^2} \cos(\Omega_0 t)
\end{eqnarray}
for the inversion of the two-level system. In this case the inversion oscillates periodically between its extreme values and these
oscillations are governed by the vacuum Rabi frequency
$\Omega_0 = 2|g|$.
In particular, this result demonstrates that in the case of the coupling of a two-level system to a single mode of the radiation field
the decay of the excited state of the two-level system is not described by an exponential decay law
but exhibits a periodic energy exchange which reflects the periodic exchange of a photon between the two-level system and the radiation field.

\subsection{Spontaneous emission of a photon in free space\label{free}}
If a material two-level system interacts with the quantized radiation field in free space its dynamics differs significantly
from the single-mode case described previously.
In order to exemplify these significant differences let us assume that initially the two-level system is prepared in
its excited state $|e\rangle$ and that the electromagnetic field is in its ground (vacuum) state $|0\rangle$. In the absence of any mode-selecting cavity all modes of the electromagnetic radiation field couple to this initially excited two-level system. In the Schr\"odinger picture the field operators of the electric and magnetic field strengths
are given by \cite{QED} 
\begin{eqnarray}
\hat{{\bf E}}({\bf x}) &=&\sum_{{\bf k}, i}\sqrt{\frac{\hbar \omega_{{\bf k}}}{2\epsilon_0}} 
\{i{\bf u}_{{\bf k},i}({\bf x}) \hat{a}_{{\bf k},i} - 
i{\bf u}^*_{{\bf k},i}({\bf x}) \hat{a}^{\dagger}_{{\bf k},i}\} = 
\hat{{\bf E}}_{+}({\bf x}) +  
\hat{{\bf E}}_{-}({\bf x}),\label{E}\\
\hat{{\bf B}}({\bf x}) &=&
\sum_{{\bf k}, i}\sqrt{\frac{\hbar}{2\epsilon_0\omega_{{\bf k}}}}
\{(\nabla\wedge {\bf u}_{{\bf k},i})({\bf x}) \hat{a}_{{\bf k},i} + 
(\nabla \wedge {\bf u}^*_{{\bf k},i})({\bf x}) \hat{a}^{\dagger}_{{\bf k},i}\} = \hat{{\bf B}}_{+}({\bf x}) + \hat{{\bf B}}_{-}({\bf x})\nonumber  
\end{eqnarray}
with their corresponding 'positive-' and 'negative-frequency' parts $\hat{{\bf E}}_{+}({\bf x})$ and $\hat{{\bf E}}_{-}({\bf x})$ (
 $\hat{{\bf B}}_{+}({\bf x})$ and $\hat{{\bf B}}_{-}({\bf x})$). 
If we consider an electromagnetic field in a cubic quantization volume of length $L$ and volume $V=L^3$
and impose
periodic boundary conditions a complete set of
orthonormal mode functions can be chosen in the form
\begin{eqnarray}
{\bf u}_{{\bf k}, i}({\bf x}) &=&{\bf e}_i({\bf k}) \frac{e^{i{\bf k}\cdot {\bf x}}}{\sqrt{V}},~i=1,2
\end{eqnarray}
with the wave vectors ${\bf k} = 2\pi {\bf n}/L$ (${\bf n} \in {\mathbb Z}^3)$ and with the corresponding real-valued unit-polarization vectors
${\bf e}_i({\bf k})$ (${\bf e}_1({\bf k})\cdot {\bf e}_2({\bf k}) =0$, ${\bf e}_1({\bf k})\wedge{\bf  e}_2({\bf k}) = {\bf k}/|{\bf k}|$) 
fulfilling the transversality condition ${\bf e}_i({\bf k}) \cdot {\bf k} =0$ for $i=1,2$.
Within a perturbative treatment of the spontaneous photon emission process
according to Fermi's Golden rule
the resulting decay of the two-level system is governed by the rate \cite{QED}
\begin{eqnarray}
\Gamma &=& \sum_{{\bf k},i} \frac{2\pi}{\hbar}\mid
\langle e|\hat{{\bf d}}|g\rangle \cdot \hat{{\bf E}}_{+}({\bf x}_0)
\mid^2 \delta (E_g + \hbar \omega_{{\bf k}} - E_e) =
\frac{\omega_{eg}^3 |\langle e| \hat{{\bf d}}|g\rangle |^2}{3\pi \epsilon_0 \hbar c^3}\nonumber\\
&&
\label{spontdec}
\end{eqnarray}
with the transition frequency $\omega_{eg} =\omega_{eg}$.
Within the framework of the dipole approximation
${\bf x}_0$ denotes the position of the center-of-mass of the two-level system.
It is apparent from Eq.(\ref{spontdec}) that the spontaneous decay rate depends on the number of field modes per unit energy
which couple resonantly, i.e. with $\omega_{{\bf k}} = (E_e - E_g)/\hbar$, to the spontaneously decaying two-level system.
Thus, altering the mode structure of these relevant field modes by a cavity, for example,
modifies the spontaneous decay rate.
This effects was confirmed in experiments with Rydberg
atoms of sodium \cite{Haroche83} excited in a niobium superconducting
cavity.
Rydberg
atoms are especially suitable for the experimental verification of
modifications of
spontaneous decay rates since they posses large 
dipole matrix element on micro-wave transitions. 
Therefore, the spontaneous
emission rate for the atomic transition 23S$\to$22P was investigated which was in resonance
with the niobium superconducting cavity at $340$GHz. 
Thus, the free-space value of
$\Gamma=150\mathrm{s}^{-1}$ could be increased to a value of $\Gamma_{cavity}=8\cdot
10^4\mathrm{s}^{-1}$.

A non-perturbative treatment of
the exponential decay of an excited two-level system has already been given  by Weisskopf and Wigner
\cite{WignerWeisskopf1,WignerWeisskopf2}
in their seminal work. If the initially prepared quantum state of the matter-field system is of the form
$|e\rangle\otimes |0\rangle$ the time evolution of the resulting quantum state can be decomposed according to
\begin{eqnarray}
|\psi\rangle_t &=& \sum_{{\bf k}, i}
a_{g {\bf k} i}(t) e^{-i(E_g + \hbar \omega_{{\bf k}}) t/\hbar}|g\rangle \otimes \hat{a}^{\dagger}_{{\bf k},i}|0\rangle    
+
a_{e}(t) e^{-iE_e t/\hbar}|e\rangle \otimes |0\rangle.
\end{eqnarray}
The resulting dynamics is described by the time dependent Schr\"odinger equation with Hamiltonian
\begin{eqnarray}
\hat{H} &=&   E_g|g\rangle \langle g| + E_e |e\rangle \langle e| + 
\sum_{{\bf k}\in I,i} \hbar \omega_{{\bf k}} \hat{a}^{\dagger}_{{\bf k},i} \hat{a}_{{\bf k},i}
-\nonumber\\
&&|e\rangle \langle g|~\langle e|\hat{{\bf d}}|g\rangle\cdot \hat{{\bf E}}_{+}({\bf x}_0)  - 
|g\rangle \langle e|~\langle e|\hat{{\bf d}}|g\rangle^*\cdot \hat{{\bf E}}_{-}({\bf x}_0) 
\label{Hint} 
\end{eqnarray}
in the dipole- and rotating-wave approximation \cite{QED}. The rotating wave approximation takes into
account the interaction of the two-level system
with the almost
resonant field modes (${\bf k}\in I$)   
within a frequency interval of width
$\Delta \omega \gg \Gamma$ centered around the transition frequency $\omega_{eg}$ in a non-perturbative way.
The influence of all other (non-resonant) field modes can be taken into account perturbatively. In particular, 
in second order perturbation theory these non-resonant modes give rise to a Lamb shift \cite{Lambshift1,Lambshift2}.
It should be mentioned that, contrary to the almost resonant modes, for a consistent treatment of this energy shift the coupling
of the non-resonant modes of the radiation field to the two-level atom cannot be treated in the dipole approximation
because also high-frequency field modes have to be taken into account.
Furthermore, mass renormalization has to be included.
This way a well-defined and finite expression for the Lamb shift can even be obtained within a
non-relativistic quantum electrodynamical description \cite{Lambshift2}. 

The time dependent Schr\"odinger equation is equivalent to the differential equations 
\begin{eqnarray}
\dot{a}_e(t) &=& -\frac{\langle e|\hat{{\bf d}}|g\rangle}{\hbar}
\sum_{{\bf k} \in I,i}  \sqrt{\frac{\hbar \omega_{{\bf k}}}{2\epsilon_0}}
{\bf u}_{{\bf k},i}({\bf x}_0)
a_{g {\bf k} i}(t) e^{i(E_e - E_g - \hbar\omega_{{\bf k}})t/\hbar},\nonumber\\
\dot{a}_{g {\bf k} i}(t) &=& \frac{\langle e|\hat{{\bf d}}|g\rangle^*}{\hbar}  \sqrt{\frac{\hbar \omega_{{\bf k}}}{2\epsilon_0}}
{\bf u}_{{\bf k},i}^*({\bf x}_0) a_{e}(t) e^{-i(E_e - E_g - \hbar\omega_{{\bf k}})t/\hbar}
\label{Wigner}
\end{eqnarray}
for the probability amplitudes which yield the 
integro-differential equation
\begin{eqnarray}
a_e(t) &=&  -
\sum_{{\bf k} \in I,i}
{\frac{\hbar \omega_{{\bf k}}| \langle e|\hat{{\bf d}}|g\rangle \cdot {\bf u}_{{\bf k},i}({\bf x}_0)|^2}{2\hbar^2\epsilon_0}}
\int_0^t dt'~ a_e(t') e^{i(E_e - E_g - \hbar\omega_{{\bf k}})(t-t')/\hbar}  
\label{integro}
\end{eqnarray}
for the probability amplitude $a_e(t)$ of observing the two-level system in its excited state $|e\rangle$ at time $t>0$.
In the continuum limit of a large
quantization volume $V$ the summation over the modes of the electromagnetic field can be approximated by
\begin{eqnarray}
\sum_{{\bf k} \in I, i} &\longrightarrow& \sum_{i}\frac{V}{(2\pi)^3}\int_{{\bf k}\in I} d^3{\bf k} =
\sum_{i}\int_{\omega \in I} d\omega \rho(\omega)
\end{eqnarray} 
with $\rho(\omega) = 4\pi V\omega^2/(8\pi^3 c^3)$ enumerating the number of field modes per frequency and per polarization.
As long as the spontaneous decay rate $\Gamma$ of Eq.(\ref{spontdec}) is much smaller than the resonance frequency
$\omega_{eg}$ and the frequency range over which
the integrand of Eq.(\ref{integro}) varies significantly, the 'pole approximation' \cite{WignerWeisskopf1,WignerWeisskopf2,pole} may by applied so that Eq.(\ref{integro}) reduces to
\begin{eqnarray}
a_e(t)&=& - \int_0^t dt'~\Gamma  a_e(t')\delta (t-t') = -\frac{\Gamma}{2}a_e(t).
\label{ae}
\end{eqnarray}
Within this approximation
the initially excited two-level atom decays exponentially with rate (\ref{spontdec}) so that for  $t>0$ the quantum state
is given by
\begin{eqnarray}
|\psi\rangle_t &=&  \sum_{{\bf k}\in I, i} (-i) \langle e|\hat{{\bf d}}|g\rangle^* \cdot {\bf u}^*_{{\bf k}, i}({\bf x}_0)
\sqrt{\frac{\hbar \omega_{{\bf k}}}{2\epsilon_0}}
\frac{e^{-i(E_g + \hbar \omega_{{\bf k}})t/\hbar} - e^{-i(E_e -i\hbar\Gamma/2)t/\hbar}}
{E_e - E_g - \hbar \omega_{{\bf k}} - i\hbar \Gamma/2} \hat{a}^{\dagger}_{{\bf k},i}|0\rangle\otimes |g\rangle +\nonumber\\
&&e^{-i(E_e - i\hbar\Gamma/2)t/\hbar}|e\rangle\otimes |0\rangle.
\label{1-photon}
\end{eqnarray}
%
Thus, at the end of the spontaneous decay process the two-level system
approaches a separable pure quantum state with the two-level system in its lower energy eigenstate $|g\rangle$.
In this quantum state the mean electric and magnetic field strengths vanish so that
the (normally ordered) energy density of the electromagnetic field provides a local
measure for the statistical uncertainty of these electromagnetic field strengths.

According to Eq.(\ref{1-photon}) after the completion of the photon emission process 
both the two-level system and the electromagnetic field are in pure states. In view of this
asymptotic separation it is of interest to ask whether a one-photon state of the electromagnetic field exists 
which is capable of exciting a
material system, such as an 
atom, perfectly in free space from an initially prepared ground state $|g\rangle$ to an excited state $|e\rangle$ by photon
absorption. Within our quantum electrodynamical model this question can be answered in the affirmative
in a straight forward way . For this purpose one has to solve Eqs.(\ref{Wigner}) subject to the final-state
condition that at a particular time, say $t=0$, the two-level system is in its excited state and the radiation field in its vacuum state.
It is straight forward to show that for $t\leq 0$ this advanced solution of Eqs.(\ref{Wigner}) is given by the quantum state
\begin{eqnarray}
|\psi\rangle_t &=&  \sum_{{\bf k}\in I, i} (-i) \langle e|\hat{{\bf d}}|g\rangle^* \cdot {\bf u}^*_{{\bf k}, i}({\bf x}_0)
\sqrt{\frac{\hbar \omega_{{\bf k}}}{2\epsilon_0}}
\frac{e^{-i(E_g + \hbar \omega_{{\bf k}})t/\hbar} - e^{-i(E_e + i\hbar\Gamma/2)t/\hbar}}
{E_e - E_g - \hbar \omega_{{\bf k}} + i\hbar \Gamma/2} \hat{a}^{\dagger}_{{\bf k},i}|0\rangle\otimes |g\rangle +\nonumber\\
&&e^{-i(E_e + i\hbar\Gamma/2)t/\hbar}|e\rangle\otimes |0\rangle.
\label{out-photon}
\end{eqnarray}
Indeed,
for $t \to -\infty$ this quantum state is separable. It describes the two-level system initially prepared in state $|g\rangle$
with the radiation field prepared in the particular pure one-photon state which finally excites the two-level system
to its excited state $|e\rangle$ perfectly by absorption of a single photon.
For times $t>0$ the continuation of this time evolution is described by the quantum state of Eq.(\ref{1-photon}).

\section{Quantum electrodynamics with controlled mode selection\label{III}}

In this section characteristic quantum phenomena originating from the exchange of energy between a two-level system and
a single-photon radiation field inside a cavity are discussed. 
Firstly, we consider a possibly large but closed spherical cavity
which may contain
many almost resonant field modes coupling to a material two-level system positioned in the center of the cavity. 
By varying the size of this cavity it is possible
to describe in a uniform way the transition between the extreme cases of 
coupling to a single mode of the radiation field and of the free-space limit \cite{Alber1992,AN}.
Secondly, we discuss
the dynamics of spontaneous photon emission by a two-level system in a half-open parabolic cavity. 

\subsection{Photon exchange in a closed spherical cavity}

The dynamics of a material two-level system positioned at the center ${\bf x}_0$ of a spherical cavity,
which supports almost resonant field modes, can be described within the theoretical framework discussed in Sec.\ref{free}.
In the dipole- and rotating wave approximation again the dynamics of this quantum system can be described by the Hamiltonian of
Eq.(\ref{Hint}) in the Schr\"odering picture. However, now the electric field operator of Eq.(\ref{E}) has to be constructed
with the help of mode functions ${\bf u}_l({\bf x})$ which are appropriate for a spherical cavity. For this purpose we
solve the Helmholtz equation (\ref{Helmholtz}) with the boundary condition of an ideal metallic spherical cavity.
Thus, the tangential component of ${\bf u}_l({\bf x})$ and the normal component of $(\nabla\wedge {\bf u}_l)({\bf x})$ have to
vanish at the boundary of the spherical cavity. The corresponding solutions of the Helmholtz equation determine the relevant
set of possible discrete eigenfrequencies $\omega_l$ inside this cavity. For a spherical cavity of radius $R$
the possible mode functions are of the form \cite{AN}
\begin{eqnarray}
{\bf U}_{nLM}({\bf x}) &=& N_{nL} j_L(\omega_{nL}r/c) {\bf X}_{LM}(({\bf x}-{\bf x}_0)/|{\bf x}-{\bf x}_0|),\nonumber\\
{\bf V}_{nLM}({\bf x}) &=& N_{nL} \frac{ic}{\omega_{nL}} (\nabla \wedge j_L)(\omega_{nL}r/c)
{\bf X}_{LM}(({\bf x}-{\bf x}_0)/|{\bf x}-{\bf x}_0|)
\end{eqnarray}
with the mode index $l \equiv (nLM)$, the vector spherical harmonics \cite{Jackson}
\begin{eqnarray}
{\bf X}_{LM}({\bf y}/|{\bf y}|) &=& -\frac{i}{\sqrt{L(L+1)}}{\bf y}\wedge
(\nabla Y_L^M)({\bf y}/|{\bf y}|),
\end{eqnarray}
and with the spherical harmonics $Y_L^M({\bf y}/|{\bf y}|)$
($L\in {\mathbb N}_0, -L\leq M \in {\mathbb Z}\leq M$)\cite{harmonics}.
The dependence of these mode functions on the radial coordinate $r=|{\bf x}-{\bf x}_0|$
is determined by the regular spherical Bessel functions \cite{harmonics}
whose asymptotic behavior is given by
\begin{eqnarray}
&&\frac{x^L}{(2L+1)!!}~_{\overleftarrow{x~\ll~1}}~j_L(x)~_{\overrightarrow{x~\gg~1}}~\frac{\sin(x-L\pi/2)}{x}
\end{eqnarray}
with $(2L+1)!! = (2L+1)(2L-1)(2L-3)\cdots 3\cdot 1$.
The normalization constants $N_{nL}$
are given by
\begin{eqnarray}
N_{nL} &=&\left(\int_0^R dr r^2 j_L^2(\omega_{nL}r/c)\right)^{-1/2}~_{\overrightarrow{n~\gg~1}}~ \frac{\omega_{nL}}{c}\sqrt{\frac{2}{R}}.
\end{eqnarray}
The eigenvalues $\omega_{nL}$ of the electromagnetic field modes are determined by 
the metallic boundary conditions, i.e.
\begin{eqnarray}
j_{L}(\omega_{nL}R/c) &=&0,~~\frac{d(xj_L(x))}{dx}\mid_{x=\omega_{nL}R/c} = 0.
\end{eqnarray}
Thus, highly excited field modes with $\omega_{nL} R/c \gg 1$ are characterized by the eigenvalues
\begin{eqnarray}
\omega_{nL}R/c ~_{\overrightarrow{\omega_{nL}R/c~\gg~1}}~ \pi n + (L+1)\pi/2.
\end{eqnarray}
Note that the energy density of highly excited field modes is given by $dn/d(\hbar \omega_{nL}) = R/(\pi\hbar c)$ and is thus frequency independent.
It should also be mentioned that at the center of the spherical cavity, i.e. at ${\bf x}_0$, only the mode functions
${\bf V}_{n L=1 M=0}({\bf x})$ are non vanishing. Therefore, in the dipole approximation the two-level system
positioned at the center of the spherical cavity
can only couple to these field modes.

Inserting the relevant mode functions 
into the electric field operator of Eq.(\ref{E}) and solving the time-dependent Schr\"odinger equation with Hamiltonian
(\ref{Hint}) yields the time evolution of the quantum state $|\psi\rangle_t$ \cite{Alber1992}. As a result, under the
condition $|\psi\rangle_{t=0} = |e\rangle \otimes |0\rangle$ in the limit of an infinitely
large cavity again the results of Eqs.(\ref{1-photon})and (\ref{out-photon}) are obtained. In particular, with the help of the relevant mode functions ${\bf V}_{n L=1 M=0}({\bf x})$ the energy distribution of the resulting one-photon state can be determined in a straight forward way. In the radiation zone of the two-level system, i.e. at distances $|{\bf x} - {\bf x}_0| = r \gg c/\omega_{eg}$,  one can use the
asymptotic form of the relevant spherical Bessel functions. Thus, for the quantum state of Eqs.(\ref{1-photon}) and (\ref{out-photon})
in this region of space the energy density of the radiation field
is well approximated by
\begin{eqnarray}
_t\langle \psi|:\frac{\epsilon_0}{2}\left(\hat{{\bf E}}^2({\bf x}) + c^2 \hat{{\bf B}}({\bf x})\right):|\psi\rangle_t &=& 
\frac{3 \Gamma \hbar \omega_{eg}}{8\pi c}\frac{\sin^2\theta}{r^2} e^{-\Gamma(|t|-r/c)}\Theta(|t| -r/c).
\label{energy}
\end{eqnarray}
Thereby, $::$ denotes the normal ordering \cite{QED} of the field operators,  $\theta$ is the angle between the dipole moment $\langle e|\hat{{\bf d}}|g\rangle$ and the direction of observation $({\bf x} - {\bf x}_0)$, and $\Theta(x)$ denotes the Heaviside unit-step function with $\Theta(x) = 1$ for $x\geq 0$ and zero elsewhere.
The electromagnetic energy density of Eq.(\ref{energy})
is a local measure for the uncertainties of the electric and magnetic field strengths.
It is apparent that this energy density has an exponential shape with a spatial extension of the order of $c/\Gamma$ and
with a sudden decrease at distance $|{\bf x} - {\bf x}_0| = r = c|t|$ from the two-level system.
Due to energy conservation the total field energy contained in this one-photon wave packet equals
$ \hbar \omega_{eg}(1-e^{-\Gamma |t|})$. The mean field energy density can also be decomposed into
its electric and magnetic contributions (which are equal) and into its various polarization components. Thus, 
along direction ${\bf e}$ the polarization component of the mean electric energy density, for example, can be
represented in the form
\begin{eqnarray}
_t\langle \psi|:\frac{\epsilon_0}{2}({\bf e} \cdot \hat{{\bf E}}({\bf x}))^2:|\psi\rangle_t &=& 
\mid {\bf e} \cdot {\cal E}({\bf x}, t)\mid^2
\label{energy1}
\end{eqnarray}
with the complex-valued electric one-photon energy-density amplitude
\begin{eqnarray}
{\cal E}({\bf x}, t) &=&
-i \sqrt{\frac{3\Gamma \hbar \omega_{eg}}{16\pi c}}\Theta(|t|-r/c) e^{i{\rm sgn}(t)\omega_{eg} (|t|-r/c)}
e^{-\Gamma(|t|-r/c)/2}
\frac{\sin\theta}{r}{\bf e}_{\theta}
\label{eampl}
\end{eqnarray}
and with ${\rm sgn}(t)$ denoting the sign of $t$.
Eq.(\ref{eampl}) is valid in the radiation zone, i.e. for $\omega_{eg} r/c \gg 1$, as long as $\omega_{eg}\gg \Gamma/2$
(compare with Eq.(\ref{ae}) and the validity of the pole approximation).
It demonstrates that in the radiation zone the electromagnetic field energy is concentrated completely in
the polarization direction ${\bf e}_{\theta}$.
Integrating Eq.(\ref{energy1}) over all space (thereby neglecting contributions outside the radiation zone)
the electric field energy at time $t$ is given by
\begin{eqnarray}
\int_{{\mathbb R}^3} d^3{\bf x}~\mid {\cal E}({\bf x}, t)\mid^2 = \frac{\hbar\omega_{eg}}{2}(1-e^{-\Gamma |t|}).
\label{elen}
\end{eqnarray}
The analogous expression for the magnetic one-photon energy-density amplitude can be obtained from Eq.(\ref{eampl}) by the replacement ${\bf e}_{\theta} ~\longrightarrow~ {\bf e}_{\varphi}$. Thus, the magnetic energy density is concentrated completely 
in the direction ${\bf e}_{\varphi}$ and
its integrated field energy contribution is also given by Eq.(\ref{elen}).

\begin{figure}
\begin{center}
      \scalebox{0.4}{\includegraphics{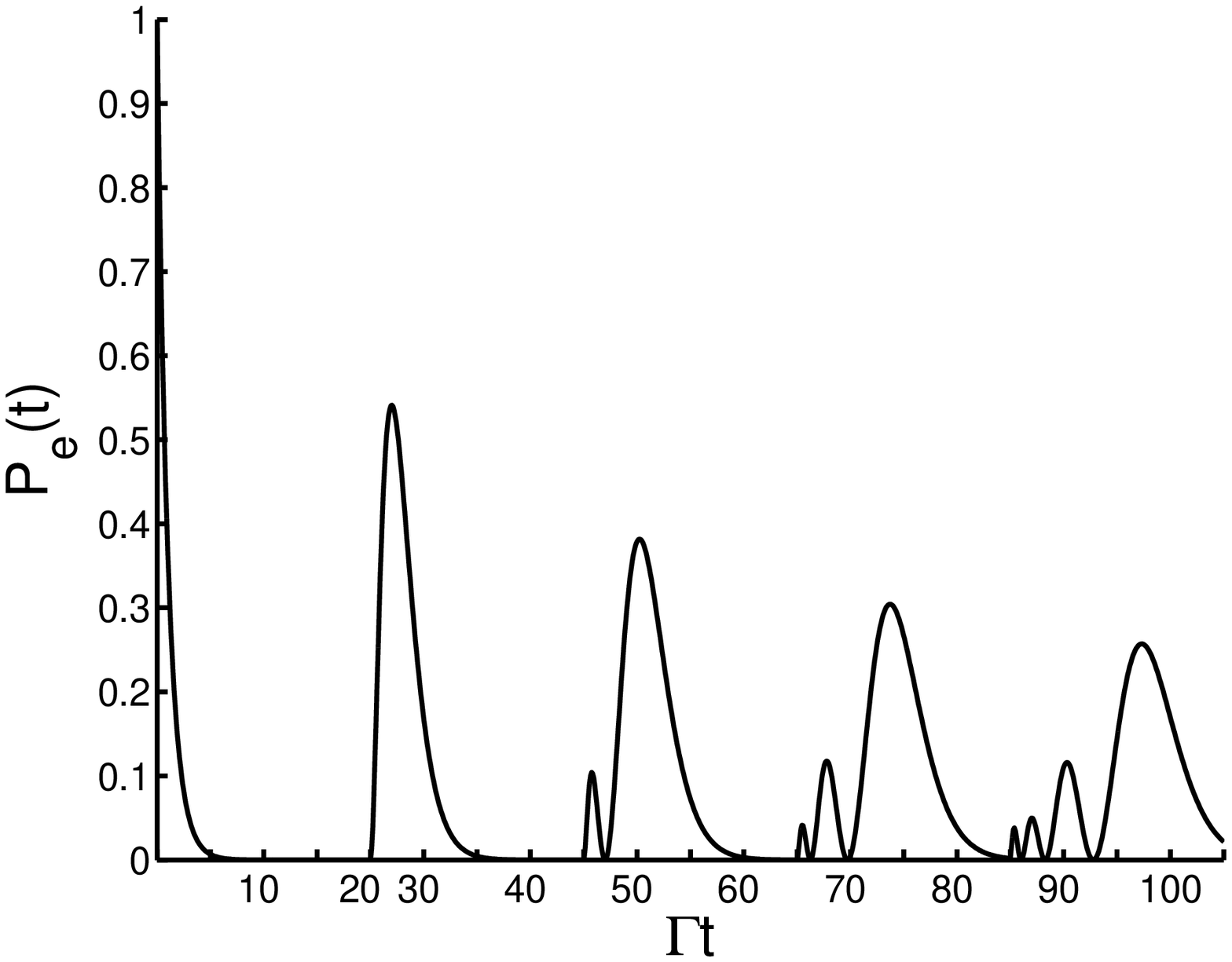}}
      \scalebox{0.4}{\includegraphics{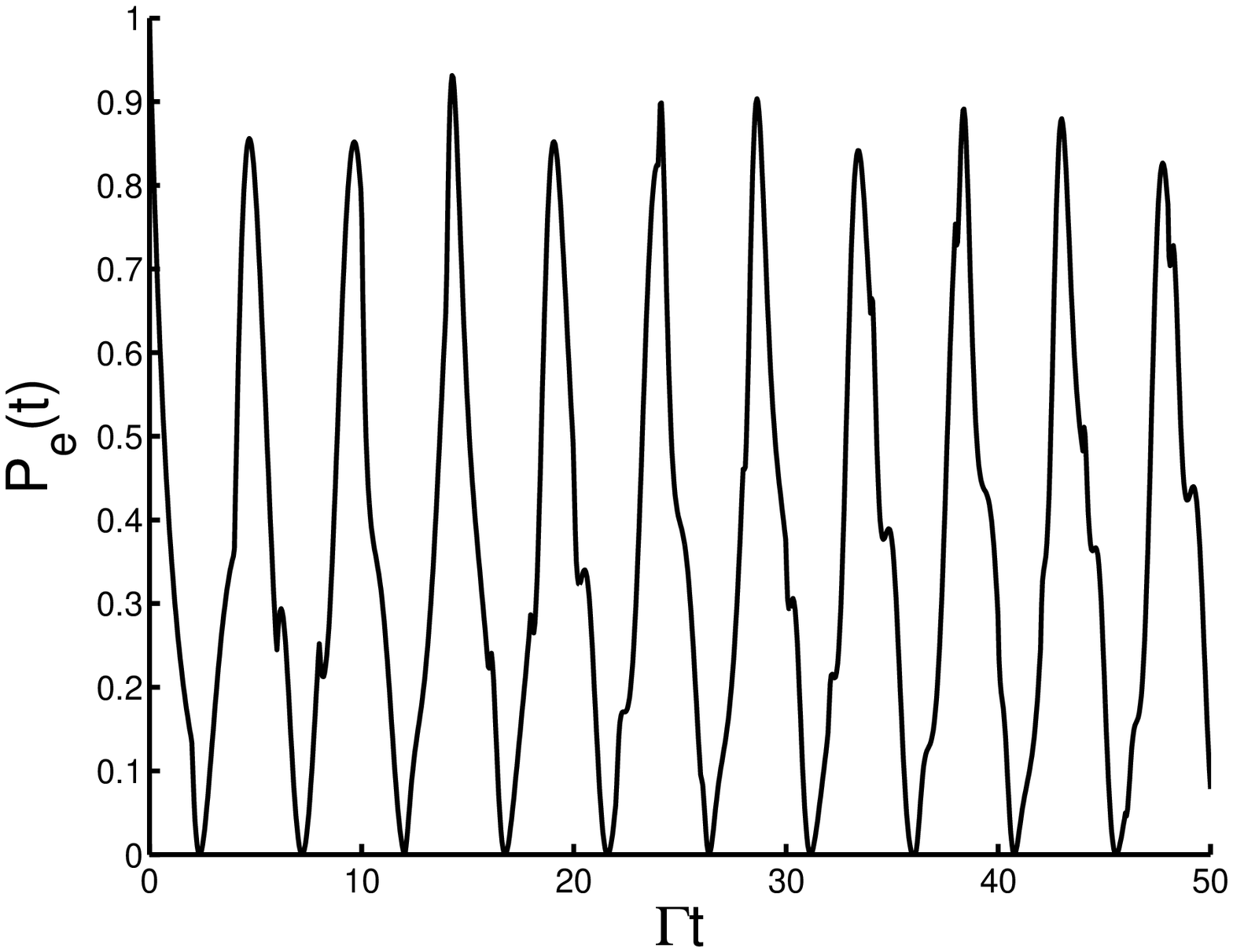}}
  \end{center}
  \caption{Time dependence of the probability $P_e(t)$ of observing the spontaneously decaying two-level system in its excited state at the center of a closed spherical cavity: The number of resonantly interacting field modes is of the order of $\Gamma R/c$ and depends
  on the size of the cavity $R$. For $\Gamma R/c =10$ (left figure) a spatially localized photon wave packet is generated by spontaneous emission
  and can be reabsorbed again by the two-level system at the center of the cavity at later times. 
  For $\Gamma R/c =1$ (right figure) only a small number of
  cavity modes interact resonantly and the two-level system performs approximate Rabi oscillations governed by the vacuum Rabi frequency.}
  \label{dyn}
\end{figure}

For finite values of the radius of the spherical cavity $R$ the dynamics of the photon exchange with the two-level system changes significantly.
In the extreme limit of a small cavity in which only one cavity mode is in resonance with the spontaneously decaying two-level system the dynamics
reduces to the results of the Jaynes-Cummings-Paul model discussed previously. In cases in which this resonant field mode is highly excited this
single-mode limit is realized if 
$\Gamma dn/d(\omega_{n L=1}) = \Gamma R/(\pi c)\ll 1$ so that the cavity is small
in comparison with the extension of a one-photon wave packet which is generated by spontaneous emission in free space and whose
spatial extension would be of the order of $c/\Gamma$. For larger values of the cavity radius $R$ the two-level system starts
to couple to more and more field modes almost resonantly so that eventually in the infinite-cavity limit the free-field dynamics is approached. 
It can be shown that for almost resonant coupling of the two-level system to highly excited modes of the spherical cavity, i.e.
$\omega_{eg}R/c\gg 1$, for $t\geq 0$ the time evolution of the excited-state probability amplitude  for an initially excited two-level system is given by
\begin{eqnarray}
\langle e|\otimes \langle 0|\psi\rangle_t &=&
e^{-i(E_e - i\hbar\Gamma/2)t/\hbar} + \sum_{M=1}^{\infty}\Theta(t-2MR/c)e^{-i(E_e - i\hbar \Gamma/2)(t - 2MR/c)/\hbar}\times\nonumber\\
&&\left\{
\sum_{r=0}^{M-1}  {M-1 \choose r}
\frac{[-\Gamma(t-2MR/c)]^{1+r}}{(1+r)!}\right\}
\end{eqnarray}
In Fig.\ref{dyn} this time evolution is depicted for various sizes of the spherical cavity.
For small cavities the characteristic almost periodic energy exchange between the two-level system and the radiation field is apparent. For larger
cavity sizes this dynamics is modified. For large cavities with $\Gamma dn/d(\omega_{n L=1}) = \Gamma R/(\pi c)\gg 1$ the initially excited
two-level system decays approximately exponentially with rate $\Gamma$ and is excited again at later times
by the spontaneously generated one-photon wave packet whenever it returns again to
the center of the cavity where the two-level system is located. 

\subsection{Spontaneous photon emission in a half-open parabolic cavity}

Efficient atom-light interaction in free space
\cite{Gerhardt,Sandoghdar,Tey,StobAlLeu} may provide us with less
technologically demanding solutions for quantum communication over
large distances than typical cavity quantum electrodynamical solutions \cite{CQED}.
Thus, it has gained a lot of interest recently. An
intermediate case between the small-cavity limit and the free-space limit
can be realized by a large closed cavity 
\cite{Alber1992} in which an atom couples to a large number of modes and which has been discussed in the previous section.
A half-cavity, i.e. a cavity with one mirror, constitutes another example
of such a case. The structure of standing light waves in front of a
mirror was already analyzed in the pioneering work of
K. Drexhage~\cite{Drexhage1970}. Later on, it has been 
verified experimentally that under such circumstances one can witness a change of the density
of states of the 
electromagnetic field modes near the atom by measuring its spontaneous emission rate
\cite{Eschner2001}.  A simplified one-dimensional scalar model of a
laser-driven atom in a half-cavity has been discussed
in~\cite{Dorner2002}, for example.

\begin{figure}
\begin{center}
      \scalebox{0.6}{\includegraphics{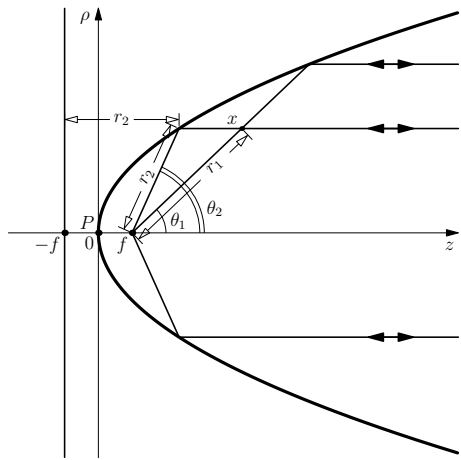}}
  \end{center}
  \caption{Spontaneous photon emission in a parabolic cavity: The two-level system is positioned in 
  the focus $F$ of a metallic parabola with focal length $f$. All light rays emanating from $F$ which are reflected at the parabolic boundary
  leave the cavity by propagating parallel
  to the symmetry axis. They all accumulate the same phase (eikonal) of magnitude $\omega_{eg}(z+f)/c$ which is the same as if these light rays had
  started in phase from the plane $z=-f$. There are always two possible trajectories to any point ${\bf x }$ inside the cavity. In the
  semiclassical limit, i.e. $f\gg c/\omega_{eg}$, these two classes of trajectories give rise
  to the spherical-wave and the plane-wave contributions to the complex-valued energy density amplitude of Eq.(\ref{eampl1}).}
  \label{parabola}
\end{figure}

As a second example for modifications of the quantum electrodynamical interaction between matter and the radiation field
originating from controlled mode engineering in the following we discuss
the spontaneous decay of a two-level system, such as an ion \cite{Sondermann2007},
in a half-open cavity with a parabolic shape.
The two-level system is assumed to be positioned in the focus $F$ of an axially symmetric parabola whose
boundary is formed by an ideal metal and is described by the equation $z = \rho^2/(4f)$ (compare with Fig.\ref{parabola}).
The coordinate
$z$ measures distances
from point $P$ along the symmetry axis and $\rho$ denotes distances perpendicular to the symmetry axis.  
The focal point $F$ of the parabola has coordinates $(z=f, \rho =0)$ with $f>0$ denoting the focal length.
A defining property of any parabola is the fact that the distance between any point on its surface 
and the focal point $F$ equals the 
distance to the plane perpendicular to the symmetry axis located at $z=-f$. 
  
We are particularly interested in  
possible changes of the spontaneous photon emission process of a two-level system
resulting from the presence of the parabolic boundary conditions.
The parabolic shape of the mirror ensures that, if a light beam is sent
parallel to the symmetry axis of the parabola towards the two-level system, this two-level system
interacts with the
light coming from the whole $4 \pi$ solid angle.
Similarly, 
the whole light resulting from spontaneous
decay of the two-level system is redirected into a beam propagating parallel to the symmetry axis.
Thinking in terms of a semiclassical ray-picture only light rays which are emitted from the two-level
system along the (negative part of the) symmetry axis return again to the two-level system. Contrary to
the closed spherical cavity discussed in the previous section all other
emitted light rays leave the cavity without any reexcitation of the two-level system. 
Thus, if the dipole moment of the two-level system is oriented along the symmetry axis photon emission along
the symmetry axis is suppressed and we do not expect any significant modification of the spontaneous decay process
of the two-level system due to the presence of the cavity in cases in which the focal length $f$ is large in comparison
with the wavelength of a spontaneously emitted photon \cite{Stobinska-Alicki}.
Nevertheless, 
in analogy to 
a small cavity or to spontaneous emission in front of a planar mirror \cite{Morawitz1969}
significant changes of the spontaneous decay rate are expected
if 
$f$ becomes comparable to the wavelength
of a spontaneously emitted photon. 

\subsubsection{Time evolution of a photon wave packet}

First of all, let us consider the dynamics of an almost resonant photon and its resonant energy exchange with a
two-level system positioned in the focal point $F$ in the semiclassical limit
in which the focal length of the parabola is 
large in comparison with the photon's wavelength, i.e. $f \gg c/\omega_{eg}$.
Furthermore let us concentrate on cases in which the two-level system's 
dipole moment $\langle e|\hat{{\bf d}}|g\rangle$ is oriented parallel to the symmetry axis of the parabola. 
In particular, we want to investigate solutions of the time dependent Schr\"odinger equation $|\psi\rangle_t$ for which
at a certain instant of time, say $t=0$,
the two-level system is excited and the electromagnetic
radiation field is in its vacuum state, i.e. $|\psi\rangle_{t=0} = |e\rangle \otimes |0\rangle$. Thus, this case describes 
spontaneous emission of a photon for $t>0$ and perfect excitation of the two-level system by a one-photon field for $t<0$.
As $f\gg c/\omega_{eg}$,
in the radiation zone, i.e. for $\omega_{eg}|{\bf x} - {\bf x}_0|/c \gg 1$,
the  distribution of the energy density of the electromagnetic field 
can be determined with the help of semiclassical methods. This is due to the fact that the one-photon energy density amplitudes
of the electric and magnetic field energies (compare with 
Eq.(\ref{eampl})) are rapidly oscillating functions of the radial variable  $r= |{\bf x} - {\bf x}_0|$ with a slowly varying envelope
which decays on the length scale $c/\Gamma$. Stated differently, the condition $\omega_{eg} \gg \Gamma/2$ implies that the phase (eikonal)
of these amplitudes exhibits many oscillations within the region of support of these amplitudes.

In the radiation zone the one-photon energy density amplitudes are solutions of the wave equation whose semiclassical
solutions \cite{Maslov}
can be constructed with the help of the classical light rays inside the parabolic cavity. 
Thus, the general form of the energy-density amplitudes of the electric and magnetic field  can be constructed semiclassically in two
steps. Firstly, one determines their form inside a sphere of radius $R_0$ around the focal point $F\equiv {\bf x}_0$ so that
the condition $c/\omega_{eg}\ll R_0 \ll f$ is fulfilled.
For the electric field this amplitude is given by Eq.(\ref{eampl}) and for the magnetic field
it can be obtained from Eq.(\ref{eampl}) by the replacement ${\bf e}_{\theta}~\longrightarrow~{\bf e}_{\varphi}$. Semiclassically,
with this form of these amplitudes one can associate a Lagrangian manifold \cite{Maslov,Arnold} of radially outgoing
straight-line trajectories which start
at the position of the two-level system $F$ and
which propagate with the speed of light. The 
relevant polarization direction remain constant during transport along these radial trajectories. 
In a second step one determines the most general semiclassical solution of the wave equation 
within
the parabolic cavity
but
outside this sphere of radius $R_0$.
For this purpose one has to construct the light rays outside this sphere
thereby taking into account that they 
are reflected at the boundary of the cavity according to the classical reflection law (equal
incoming and outgoing angles).
Apart from the reflection process where one has to take into account
the boundary conditions of an ideal metal the polarization directions remain constant during the propagation along any of these
classical trajectories. The geometry of the parabola implies that after reflection at the metallic boundary the classical trajectories
propagate parallel to the symmetry axis. Matching the solutions inside the sphere of radius $R_0$ and outside the sphere in a smooth way
one
obtains the one-photon energy-density amplitude of the electric field at any point ${\bf x}\equiv (z,\rho)$
inside the parabolic cavity.
Whereas
for times $|t|< f/c$ this energy-density amplitude 
is given by Eq.(\ref{eampl}), for times $t>f/c$ it is modified due to reflections of light rays at the boundary of 
the parabolic cavity and assumes the form
\begin{eqnarray}
{\cal E}({\bf x}, t) &=&
-i \sqrt{\frac{3\Gamma \hbar \omega_{eg}}{16\pi c}}
\Theta(|t|-r_1(z,\rho)/c)
e^{i{\rm sgn}(t)\omega_{eg} (|t|-r_1(z,\rho)/c)}
e^{-\Gamma(|t|-r_1(z,\rho)/c)/2}
\frac{\sin \left(\theta_1 (z,\rho)\right)}{r_1(z,\rho)}
{\bf e}_{\theta_1} 
+\nonumber\\
&&
-i \sqrt{\frac{3\Gamma \hbar \omega_{eg}}{16\pi c}}
\Theta(|t|-\frac{z+f}{c})
e^{i{\rm sgn}(t)\omega_{eg} (|t|-(z+f)/c)}
e^{-\Gamma(|t|-(z+f)/c)/2}
\frac{\sin \left(\theta_2(\rho)\right)}{r_2(\rho)}
{\bf e}_{\rho} 
\label{eampl1}
\end{eqnarray}
with
\begin{eqnarray}
\sin(\theta_1(z,\rho)) =\frac{\rho}{r_1(z,\rho)}&,&r_1(z,\rho) = \sqrt{(z-f)^2 + \rho^2}\nonumber\\
\sin(\theta_2(\rho)) =\frac{\rho}{r_2(\rho)}&,&r_2(\rho) = f(1+\frac{\rho^2}{4f^2}).\nonumber
\end{eqnarray}
Therefore, at a fixed time $t$ the contribution at position $ {\bf x} \equiv(z,\rho)$ results from the interference of
contributions originating from radially propagating
direct light rays
(first term on the r.h.s. of Eq.(\ref{eampl1}))
with the contributions from light rays reflected at the boundary of the cavity 
(second term on the r.h.s. of Eq.(\ref{eampl1})).
Whereas the first type of contributions give rise to a slowly modulated
spherical wave front emanating from the position of the two-level system at
$F \equiv {\bf x}_0$, the reflected trajectories
give rise to a (rapidly varying)
plane wave propagating in the $z$-direction with a slowly varying amplitude.
This plane wave is a consequence of the parabolic geometry of the cavity and the fact that for any point on
the boundary of the cavity its distances from the focal point $F$ and from the plane $z=-f$ are equal. Therefore,
all light rays
emitted at any angle $\theta_2$ at $F$ and
reflected at the boundary of the parabolic cavity accumulate the same phase (eikonal).
This phase is the same as the one originating from a ficticious set of trajectories
 starting from the plane $z=-f$ and
propagating along the $z$-axis with the speed of light. 
For  large focal lengths in the sense that $2f \gg c/\Gamma$ the two contributions to the energy-density amplitude are well 
separated in space apart from small regions around the boundary of the parabolic cavity. As a consequence interferences
between contributions of these two different types of classical trajectories disappear. 
Furthermore, 
for a fixed value of $\rho$ 
the spherical-wave contribution to Eq.(\ref{eampl1})
becomes vanishingly small in comparison with the plane-wave contribution in the limit of large values of $z$.
It is apparent from Eq.(\ref{eampl1}) that the polarization properties of the spherical-wave contribution are not
changed by the cavity and are the same as in free space. However, the polarization features of the plane-wave contribution
are significantly different. At any point ${\bf x}$ its polarization is directed in the ${\bf e}_{\rho}$-direction
so that the vector field ${\cal E}({\bf x},t)$ represents a vortex field with respect to its polarization properties.
The singularity at the symmetry axis is suppressed by the fact that there is not coupling between the radiation field and
the two-level system along this axis because the dipole moment of the two-level system is oriented in this direction.
Finally, it should be mentioned that
the (primitive) semiclassical expression of Eq.(\ref{eampl1}) breaks down at points close to the boundary of the parabolic cavity
where transitional or uniform semiclassical approximations
\cite{Maslov,Berry} have to be employed.

\subsubsection{Modifications of the spontaneous decay rate}
\label{sec:parabolic_geometry}

In this subsection it will be demonstrated that depending on the characteristic parameters, namely the
focal length $f$, the resonant wavelength $\lambda = (2\pi)c/\omega_{eg}$,
and the orientation of the two-level system's dipole moment
$d=\langle |\hat{{\bf d}}|g\rangle$, the spontaneous decay rate of the two-level system may differ significantly from its
free-space value as given by Eq.(\ref{spontdec}). 

Since the two-level system is located in a half-open space 
we start by expanding the
electric field operator (\ref{E}) in 
mode functions suitable for the parabolic symmetry of the
problem. Following the results of \cite{BKM}
we use  mode functions of the form
\begin{equation}
{\bf E}^{\sigma}_{k,\ell,\mu}({\bf r})= \frac{k}{(2\pi)^{3/2}}
\int_{S^2} d{\bf n} \, e^{i k{\bf n} \cdot {\bf r}} \,
    {h}_{\ell,\mu}({\bf n}){\bf e}^{\sigma}({\bf n}).
\label{solution}
\end{equation}
Thereby, ${\bf k} = k{\bf n}$ denotes the wavevector, $\sigma
= 1,2$ enumerates the polarization states, and parameters $\ell = 0, \pm 1,
\pm 2, ...$ and $\mu\in(-\infty, +\infty)$ are additional mode indices. The
unit vector ${\bf n}$ and the polarization vectors ${\bf e}^{1}({\bf
  n}), {\bf e}^{2}({\bf n})$ constitute the orthonormal basis which
ensures the transversality condition
$\nabla \cdot {\bf E}=0.$
In particular, we choose these directions in the following form
\begin{eqnarray}
{\bf n} &=& (\sin\theta \cos\varphi, \sin\theta \sin\varphi,
\cos\theta),\\ {\bf e}^{1}({\bf n}) &=& (\sin\varphi,-\cos\varphi,0),
\\ {\bf e}^{2}({\bf n}) &=& (\cos\theta \cos\varphi, \cos\theta
\sin\varphi, -\sin\theta).
\end{eqnarray}
The functions
${h}_{\ell,\mu}({\bf n})$ are explicitly given by
\cite{BKM}
\begin{equation}
h_{\ell,\mu}(\theta, \varphi) =
\chi_{\mu}(\theta)\frac{e^{i\ell\varphi} }{\sqrt{ 2\pi}},
\end{equation}
with
\begin{eqnarray}
\chi_{\mu}(\theta) &=& \frac{\exp\left( -i \mu \ln[\tan \theta/2]
  \right)}{\sqrt{2\pi}\sin\theta}.
\label{solution-free}
\end{eqnarray}
One can easily check the orthogonality and completeness conditions
\begin{equation}
\int_0^{2\pi} \!\!\!\!d \varphi
\!\int_0^{\pi}\!\!\!d\theta\sin\theta\; h^*_{\ell,\mu}(\theta,
\varphi)h_{\ell',\mu'}(\theta, \varphi) = \delta_{\ell\ell'}\delta(\mu
- \mu'),
\label{ortho}
\end{equation}
\begin{equation}
\sum_{\ell=-\infty}^{+\infty}\!\!\int_{-\infty}^{+\infty}\!\!\!d\mu\;
h^*_{\ell,\mu}(\theta, \varphi)h_{\ell,\mu}(\theta', \varphi') =
\delta(\varphi - \varphi ')\frac{\delta(\theta -
  \theta')}{\sin\theta}.
\label{ortho1}
\end{equation}
Combining Eq.~(\ref{solution}) with Eq.~(\ref{ortho}) one obtains the
orthogonality of the mode functions, i.e.
\begin{equation}
\int d{\bf r} \, {\bf E}^*{}^{\sigma}_{k,\ell,\mu}({\bf r}) \cdot
{\bf E}^{\sigma'}_{k',\ell',\mu'}({\bf r}) = \delta(k-k')\delta(\mu-\mu')
\delta_{\ell\ell'}\delta_{\sigma \sigma'}.
\label{orthoelectric}
\end{equation}
If 
a two-level atom is positioned at ${\bf x}$ in free
space with the transition dipole parallel to the $z$-axis, 
in the dipole approximation the resulting spontaneous decay rate 
is given by
\begin{equation}
\Gamma ({\bf x}) = \frac{1}{(2 \pi)^2}\frac{d^2 k^3}{2\hbar
  \epsilon_0} \sum_{\sigma,\ell}\!\int_{-\infty}^{+\infty}\!\!d\mu
\int \!d{\bf n} \!\int \!d{\bf n}' f_k({\bf n},{\bf r})f^*_k({\bf
  n}',{\bf r}) \cdot e^{\sigma}_z({\bf n}) e^{\sigma}_z({\bf n}') \,
h^*_{\ell,\mu}({\bf n},k) h_{\ell,\mu}({\bf n}',k)
\label{gamma}
\end{equation}
with $k=\omega_{eg}/c$.
Furthermore, we defined
$f_k({\bf n},{\bf r}) = {\rm exp}(i k{\bf n}
  \cdot {\bf r})$, $e^{\sigma}_z({\bf n}) = {\bf e}^{\sigma}({\bf
  n}) \cdot {\bf e}_z$, and used the fact that the summation over $\ell$
produces $\delta(\varphi -\varphi')$ (see
Eq.~(\ref{ortho1})). Therefore, the relevant directions ${\bf n}$, ${\bf n}'$, and the
$z$-axis belong to the same plane which leads to the relation
\begin{equation}
\sum_{\sigma} e^{\sigma}_z({\bf n}) e^{\sigma}_z({\bf n}') =
e^{2}_z({\bf n}) e^{2}_z({\bf n}') = \sin\theta \sin\theta'.
\end{equation}
Taking into account that the integration over $\mu$ yields another
Dirac delta distribution $\delta(\theta -\theta')$ we finally obtain
\begin{equation}
\Gamma(x,y,z) \!= \!\frac{1}{(2 \pi)^2}\frac{d^2 k^3}{2\hbar
  \epsilon_0}\int^{2 \pi}_0 \!\!\!d\varphi \!\!\int^{\pi}_0 \!\!d\theta
\sin^3\theta |f_k(x,y,z;\varphi,\theta)|^2,
\label{gamma1}
\end{equation}
with ${\bf x}\equiv (x,y,z)$ and $f_k({\bf n},{\bf r})\equiv
f_k(x,y,z;\varphi,\theta)$. The relation
$|f_k(x,y,z;\varphi,\theta)|=1$ implies that we recover again the free-space result
of Eq.(\ref{spontdec}).

In the presence of a conducting parabolic mirror
the mode functions of 
Eq.~(\ref{solution})
should fulfill
the appropriate boundary
conditions of an ideal metal.
However, it is very challenging to solve
the Helmholtz equation under these boundary conditions and the transversality condition 
simultaneously.
Contrary to reference \cite{Nockel}, a simplified approximation may be obtained
by  keeping the transversality condition but relaxing the precise
conditions on the mode functions on the surface of the parabolic mirror. 
The transversality
condition relates the electric field to the geometry of the system and
therefore contributes to some geometrical factor present in the decay
rate formula.
The boundary condition ensures a possible discreteness of the
normal modes. 
As our physical system is large 
it is expected that 
slightly  changing the boundary conditions of the parabolic cavity will not
influence the decay rate significantly.

We start our approximate treatment by introducing parabolic coordinates
$(\xi, \eta, \varphi)$ which are related to Cartesian coordinates by
\begin{eqnarray}
x &=& 2\sqrt{\xi \eta} \cos\varphi, \\ y &=& 2\sqrt{\xi \eta}
\sin\varphi, \\ z &=& \xi -\eta.
\end{eqnarray}
The boundary of the parabolic mirror is described by the equation $\eta = f$.
It can be shown \cite{BKM} that
the important $\eta $-dependent part of the mode functions 
possesses the following asymptotic
behavior
\begin{equation}
 F_{\ell,\mu}(k;\eta)\sim  \frac{\cos\left\{\mu\ln2k\eta +
      k\eta- \alpha_{\ell,\mu}\right\}}{\sqrt{\eta}}
  \label{asymptotic}
\end{equation}
with a phase $\alpha_{\ell,\mu}$ whose explicit form is not important for our subsequent discussion.
Imposing the parabolic boundary condition
on Eq.~(\ref{asymptotic}) at the value $\eta = f$ results in a 
discrete set of values for $\mu_m$. 
A simple choice for these discrete values is
\begin{equation}
kf-\alpha_{\ell,\mu}=0\ ,\ \mu_m\ln 2kf = m\pi \ ,\ m=1,2,3,...
\label{periodicity}
\end{equation}
This particular choice
is consistent with the replacement of the
continuous set of modes of Eq.~(\ref{solution-free}) by the
discrete set
\begin{eqnarray}
\tilde{\chi}_{m}(\theta) &=& \frac{\sin\left(\frac{m\pi \ln[\tan
      \theta/2]}{\ln 2kf} \right)}{\sqrt{2\pi\ln
    2kf}\sin\theta}\;\mathrm{for} \; \theta\in [\theta_0 , \pi
  -\theta_0]\\ \nonumber &=& 0\ \mathrm{otherwise}
\label{solution-mirror}
\end{eqnarray}
such that
\begin{equation}
\tan\frac{\theta_0}{2}= \frac{1}{2kf}= \frac{1}{4\pi}\frac{\lambda}{f}.
\label{periodicity1}
\end{equation}
The limitation on the angle $\theta$ results from the quantization
condition (\ref{periodicity}) and from the fact that at the boundary the normal modes have
to vanish, i.e.
$\tilde{\chi}_{m}(\theta_0)=0$.  This ansatz modifies the formula for the
decay rate because the completeness condition is changed according to
\begin{equation}
\sum_m\tilde{\chi}_{m}(\theta)\tilde{\chi}_{m}(\theta')= I_{[\theta_0,
    \pi-\theta_0]}(\theta)\frac{\delta(\theta -\theta')}{\sin\theta}
\label{completeness}
\end{equation}
with $I_A$ denoting the indicator function of the set $A$. Therefore,
the integration over the angle $\theta$ involved in  Eq.~(\ref{gamma}) should be performed
over the interval $[\theta_0, \pi-\theta_0]$. This leads to
a correction of the order of $(kf)^{-4}$ which, however, is 
not
relevant for present-day experiments. 
In currently planned experiments \cite{Sondermann2007}
typical focal lengths and wavelengths are of order of $f=2$mm and 
$\lambda=250$ nm which amounts to a value of $kf\simeq 10^4$
\cite{StobAlLeu} so that $\theta_0$ is small. Therefore, we can
replace $\sin\theta$ by $\theta$. It is rather obvious that the same
is true for any reasonable choice of the boundary conditions because
the smallness of this correction is entirely due to the large value of
$kf$.

Hence the only relevant modification of the spontaneous emission rate
due to the presence of the parabolic mirror is the replacement of the plane
traveling waves $f_k({\bf n},{\bf r}) =e^{i k{\bf n} \cdot {\bf r}}$
by the standing waves
\begin{equation}
f_k({\bf n},{\bf r}) =\sqrt{2}\sin\left( k{\bf n} \cdot ({\bf r}-{\bf
  f})\right)
\label{standingwave}
\end{equation}
where the factor $\sqrt{2}$ ensures the completeness condition. 
Eq.~(\ref{standingwave}) implies that the electric field for
any mode of the form of Eq.~(\ref{solution}) vanishes at the point P of the parabola.
It leads to
the following approximate expression for the spontaneous emission rate in the presence
of a conducting parabolic mirror
\begin{equation}
\tilde{\Gamma}(x,y,z) = \frac{1}{2 \pi^2}\frac{d^2 k^3}{2\hbar
  \epsilon_0}\int^{2 \pi}_0 \!\!\!d\varphi \int^{\pi}_0 d\theta
\sin^3\theta \sin^2\{k [(x \cos\varphi + y \sin\varphi)\sin\theta +
  (z+f) \cos\theta]\}.
\label{gamma2}
\end{equation}
If the two-level system is placed at a distance from the point P which is much larger
than $\lambda = (2\pi)c/\omega_{eg}$, the interference factor $\sin^2(...)$ is averaged to
$1/2$ and the standard free-space result of Eq.~(\ref{spontdec}) is
recovered.

\begin{figure}
\begin{center}
      \scalebox{0.6}{\includegraphics{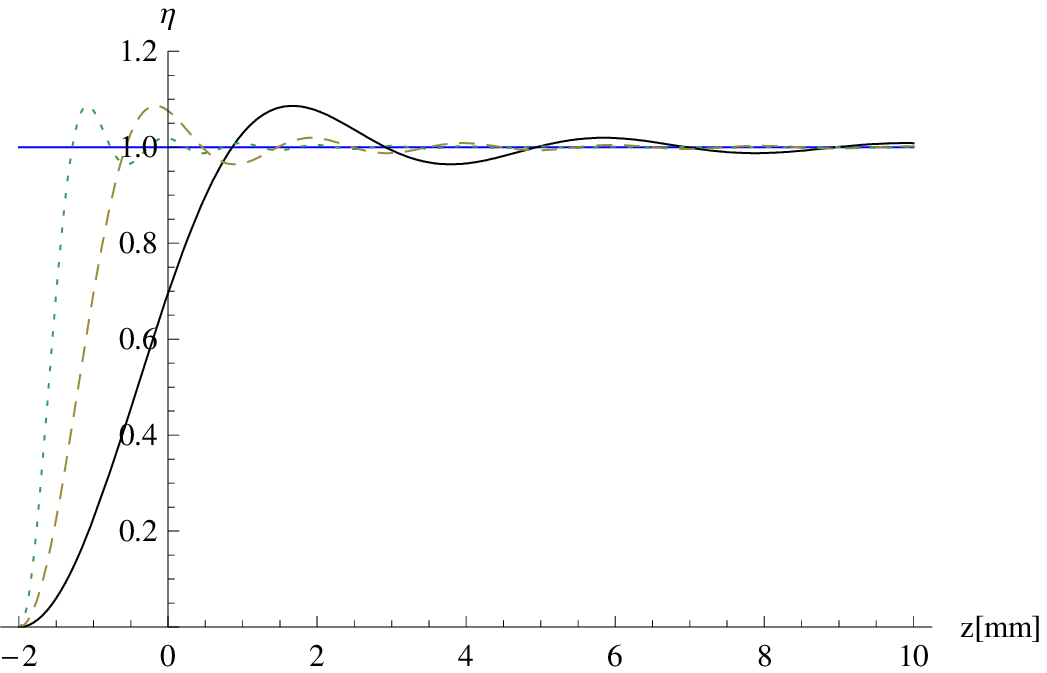}}
      \scalebox{0.6}{\includegraphics{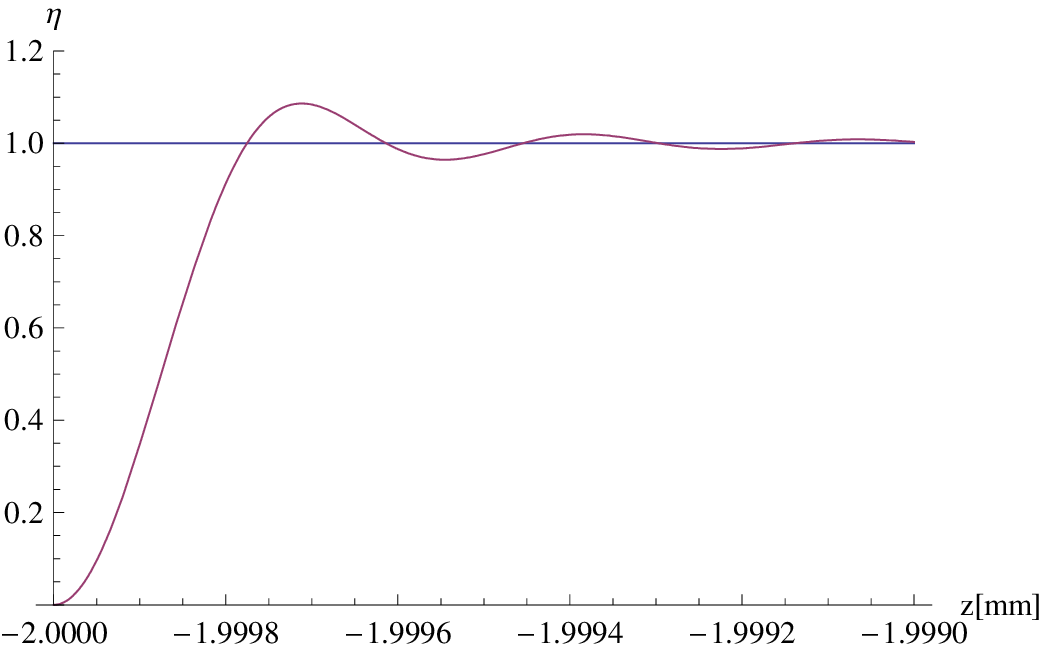}}
  \end{center}
  \caption{Modifications of the spontaneous decay rate according to Eqs.(\ref{corrections0}) and (\ref{corrections}):
    small values of $k=(2\pi)c/\omega_{eg}$ (left figure): $k=0.25\pi$
    mm$^{-1}$ -- solid line, $k=0.5\pi$ mm$^{-1}$ -- dashed
    line, $k=\pi$ mm$^{-1}$ -- dotted line; large value of
    $k=10^4$ mm$^{-1}$ (right figure).}
  \label{fig_corrections1}
\end{figure}

If the two-level system is located at the symmetry axis of the parabola, i.e. $x=y=0$, the integral in
Eq.~(\ref{gamma2}) simplifies to
\begin{equation}
\tilde{\Gamma}(z)= \eta \Gamma
\label{corrections0}
\end{equation}
with
the correction $\eta$  to the free-space decay rate $\Gamma$ being given by
\begin{equation}
\eta = \left(1+3 \frac{\cos(2k(z+f))}{4k^2(z+f)^2} - 3 \frac{\sin(2
  k(z+f))}{8k^3(z+f)^3}\right).
\label{corrections}
\end{equation}
For a focal length $f=2$ mm
this correction factor $\eta$ is depicted in
Fig.~\ref{fig_corrections1}. Its value at the focal point $F$ becomes
significant for small values of the wave vector $k = c/\omega_{eg}$. This
corresponds to cases 
in which the two-level system is close the mirror
surface, i.e. 
$|z+f|<\lambda$. 
However, for larger values of $k$ 
the variations of $\eta$ are shifted towards the mirror
surface. 
Far away from the mirror $\eta$ tends to unity
so that the decay rate reduces to its free-space value.  
According to these results modifications of the spontaneous decay rate
could be observable on a scale of $100$nm but only within
a distance of the order of the wavelength from the mirror surface.

\section{Conclusions and outlook}

Despite significant recent advances in the area of cavity quantum
electrodynamics concerning the control of photonic quantum states and
their interaction with matter in cavities, the interaction of
few-photon multi-mode quantum states with matter in free space is
still largely unexplored. A detailed understanding of this interaction
and the resulting exchange of quantum information between radiation
field and matter is not only of general physical interest but is also
necessary for promising future quantum technological applications,
such as the realization of quantum repeaters.

Here, we have discussed the simplest problem in this respect, namely
the free-space interaction of a single-photon quantum state with an
individual two-level matter system which can be realized by a trapped
atom or ion. In this elementary example it can be demonstrated
explicitly that the process of spontaneous emission of a photon is
perfectly reversible provided that one is able to control the modes
and quantum states of the radiation field in free space. For this
purpose the dynamics resulting from the mode structure of a parabolic
mirror has been discussed. It has been shown that using parabola it is
possible to perfectly convert the excitation of an appropriately
prepared asymptotically incoming plane-wave one-photon quantum state
to an atom positioned in the focus of the parabola. The resulting
dynamics depends strongly on the magnitudes of the dominantly excited
wave lengths of the one-photon state.  If these wave lengths are short
in comparison with the focal length of the parabola the propagation of
the one-photon state can be described by semiclassical methods and is
dominated by the light rays of the photon inside the parabola.  In the
opposite limit of long wave lengths diffraction effects become
important and semiclassical treatments become inappropriate.

Further experimental testing of the discussed theoretical results is
planned. In our future experiment \cite{Sondermann2007,Maiwald} we
will use a ${}^{174}\mathrm{Yb}^{2+}$ ion as a two-level system with
${}^1S_0$ and ${}^3P^0_1$ electronic levels as the ground and the
excited state respectively and no hyperfine structure.  The atomic
transition frequency $\omega_0\!\!=\!\!251.8$ nm is in the ultraviolet
regime. The ion will be trapped at the focus $f\!=\!2.1$mm of a
metallic parabolic mirror, being one electrode of a Paul trap. The rf
needle-shaped electrode will come from the back of the mirror through
a small hole.  This trap design will ensure almost full 4$\pi$ angle
of ion-light interaction in the strong focusing regime. The aberration
corrections will be done using a diffractive element located in front
of the mirror. Since the ion has only one decay channel and its dipole
moment will be parallel to the mirror axis we obtain free space
geometry. There are several methods which allow for single-photon
pulse generation with the desired spatio-temporal shape and spectral
distribution. The first relies on electro-optic modulation of a
single-photon wave packet \cite{Harris}.  Another experimentally more
accessible method applies a strongly attenuated laser pulse containing
$\overline{n}\!\!\ll \!\!1$ photons on average. This technique is
widely used in Quantum Key Distribution \cite{QKD}.  We can shape a
pulsed temporal mode electronically with modulators starting from a
continuous-wave laser.  Next we will turn it into a radially polarized
spatial doughnut mode.  After reflection from the mirror surface its
polarization, at the focal point will only contribute to polarization
parallel to the axis of the mirror and therefore will excite a linear
dipole oscillating parallel to this axis.  Using this simpler method
perfect coupling is achieved if the probability of excitation matches
the probability of finding a single photon in the pulse.  In addition,
as a third option one can generate the properly shaped single-photon
Fock state wave function conditionally using photon pairs from
parametric down conversion. This method is similar to ghost imaging in
the time domain.

Of course, none of these methods will produce an infinitely long
pulse. This is not an obstacle for our experiment however, because one
can truncate somewhat the exponential tail of the one-photon
wavepacket.  For example, truncating the pulse to a duration of five
lifetimes the excitation probability can be as high as 0.99. For
quantum-storage applications it is straightforward to expand this
scheme to a lambda transition between two long lived states
\cite{Pinotsi}. Furthermore, efficient coupling in free space opens
the possibility for non-linear optics at the single photon level.

The investigations discussed here present first steps towards the
future goal of obtaining a detailed understanding of the interaction
of individual photons with individual atoms in free space.  For future
applications in quantum information technology the exploration of the
limits of perfect exchange of quantum information between the
radiation field and these material systems is of particular interest.

\end{document}